\documentclass[twocolumn,prb,showpacs,superscriptaddress]{revtex4-1}
\usepackage{ulem}
\usepackage{graphicx}
\usepackage{amsmath} 
\usepackage{amsmath} 
\usepackage{amssymb}
\usepackage{ifthen} 
\usepackage{bm} 
\usepackage{dcolumn} 
\usepackage{color}

\newcommand{\MCtwo}{Microtechnology and Nanoscience, MC2, Chalmers University of Technology, SE-412 96 G{\"o}teborg, Sweden} 
\newcommand{\LBNL}{Molecular Foundry, Lawrence Berkeley National Laboratory, Berkeley, California 94720, U.S.A.} 
\newcommand{\Wake}{Department of Physics, Wake Forest University, Winston-Salem, North Carolina 27109, U.S.A.} 
\newcommand{\APP}{Department of Applied Physics, Chalmers University of Technology, SE-412 96 G{\"o}teborg, Sweden}
\newcommand{\ORNL}{Materials Science and Technology Division, Oak Ridge National Laboratory, Oak Ridge, Tennessee 37831-6114, U.S.A.}

\newcommand{\diff}{\mathrm{d}}

\newcommand{\self}{{\text{self}}}

\newcommand{\ACF}{{\text{ACF}}}
\newcommand{\vdWDF}{\text{vdW-DF}}

\newcommand{\nl}{\text{nl}} 
\newcommand{\Enl}{E_c^{\nl}[n]}
\newcommand{\msl}{\text{sl}} 
\newcommand{\im}{\mathrm{i}}
\newcommand{\eq}[1]{Eq.~(\ref{#1})}

\begin{document}

\title{van der Waals density functionals built upon the electron-gas tradition: \\
Facing the challenge of competing interactions}

\author{Kristian Berland}\affiliation{\MCtwo}\affiliation{\LBNL}
\author{Calvin A. Arter}\affiliation{\Wake} 
\author{Valentino R. Cooper}\affiliation{\ORNL} 
\author{Kyuho Lee}\affiliation{\LBNL}
\author{Bengt I. Lundqvist}\affiliation{\APP}
\author{Elsebeth Schr{\"o}der}\affiliation{\MCtwo} 
\author{T. Thonhauser}\affiliation{\Wake} 
\author{Per Hyldgaard}\affiliation{\MCtwo} 
\date{\today}

\begin{abstract} 
The theoretical description of sparse matter
attracts much interest, in particular for those ground-state
properties that can be described by density functional theory
(DFT). One proposed approach, the van der Waals density
functional (vdW-DF) method, rests on strong physical foundations
and offers simple yet accurate and robust functionals. 
A very recent functional within this method called
vdW-DF-cx [K. Berland and P. Hyldgaard, Phys. Rev. B 89, 035412]
stands out in 
its attempt to use an exchange energy derived from the same plasmon-based theory from which the nonlocal correlation energy was derived.
Encouraged by its good performance
for solids, layered materials, and aromatic molecules, we apply
it to several 
systems that are characterized by competing interactions.
These include the ferroelectric response in PbTiO$_3$, the adsorption of small
molecules within metal-organic frameworks (MOFs), the
graphite/diamond phase transition, and the adsorption of an aromatic-molecule
on the Ag(111) surface. Our results indicate that vdW-DF-cx is
overall well suited to tackle these challenging systems.
In addition to being a competitive density functional for sparse
matter, the vdW-DF-cx construction presents 
a more robust general purpose functional that could be applied to a range of materials problems with a variety of competing interactions.
\end{abstract}

\maketitle

\section{Introduction}

Sparse matter is important and calls for a theoretical description as accurate as one for dense matter. Despite the urgency of this need there were surprisingly few density functional theory (DFT) papers accounting for the interactions in sparse media before the 90s. 
Such interactions may be weak and of several kinds, but it is the delicate balance between them that is key to determining materials properties.
Since the 90s, there has been an upsurge in electron-structure calculations describing nonlocal correlations, in particular van der Waals (vdW) forces. A spectrum of methods have been proposed. Among them the van der Waals-density functional (vdW-DF)\cite{Dion04p246401, Thonhauser07p125112, Langreth09p084203} method stands out in its high ambitions of designing a tractable functional with strong physical foundations, thereby resulting in low computational demands and high accuracy. With this in mind, early vdW-DF method development\cite{Andersson96p102, Hult96p2029, Hult99p4708, Rydberg00p6997, Rydberg01p, Rydberg03p126402, Dion04p246401, Dion04p, Langreth05p599, Thonhauser07p125112, Langreth09p084203} focused on constructing a nonlocal correlation functional that was (i) nonempirical, that is, entirely given by the ground-state density with no external parameters, and (ii) robust, thereby creating a framework for the inclusion of dispersive or vdW interactions \cite{Eisenshitz30p491,London30p245} and, more generally, truly nonlocal correlations. 
Such a method should cover different energy scales. 
These functionals build upon the experience obtained during decades of use of their predecessors, the local density approximation\cite{Hedin71p2064, Gunnarsson76p4274, Perdew92p13244} (LDA) and the constraint-based formulations of the generalized gradient approximations (GGAs).\cite{Rasolt75p1234, Langreth75p1425, Langreth77p2884, Langreth81p446, Langreth87p497, Langreth90p175, Perdew86p8800, Perdew96p3865, Perdew96p16533}

Initially, relatively little attention was devoted to exchange. The \lq\lq{}modest\rq\rq{} goal of the pioneering vdW-DF1 functional\cite{Dion04p246401} was to describe vdW bonds with separations typical for them, 3--4 \AA, and to ensure that there was no unphysical binding from the exchange-only description.\cite{Rydberg03p126402, Dion04p246401, Langreth05p599} For these reasons, the revPBE exchange functional\cite{Zhang98p890} was chosen. Numerous applications of vdW-DF1, however, implicated  this choice of exchange as the cause of a consistent overestimation of inter-molecular separation distances and lattice constants. Subsequently, the focus turned to the development of appropriate exchange functionals to pair with the nonlocal correlation term.\cite{Thonhauser06p164106, Murray09p2754, Cooper10p161104, Klimes10p22201, Berland13p} 
One strategy was to adjust existing exchange functionals such that the total-energies of a subset of molecular systems agreed with those of advanced quantum-chemical calculations.\cite{Jurecka06p1985}

This paper sums up results obtained with a more rigorously derived exchange functional for the vdW-DF nonlocal correlation; one designed with a critical eye on consistency between exchange and correlation. This exchange functional follows a tradition of describing the fully interacting and therefore screened electron gas primarily in terms of the plasmon-response behavior.\cite{Bohr54p1,Lindhard54p1,Sawada57p372, Pines89p,Lundqvist67p193, Lundqvist67p206,Lundqvist68p117,Hedin71p2064,Langreth75p1425, Langreth77p2884} Ingredients are the adiabatic connection formula \cite{Gunnarsson76p4274, Langreth75p1425, Langreth77p2884} (ACF) and the assumption that a (single-)plasmon-pole approximation \cite{Rydberg00p6997,Rydberg01p, Dion04p246401,Hyldgaard13p} for the dielectric function, $\epsilon$, can be picked to represent the full exchange-correlation energy in the spirit of the electron-gas tradition.\cite{Lundqvist68p117, Lundqvist67p193, Lundqvist67p206,Gunnarsson76p4274,Langreth75p1425, Langreth77p2884} Further details are given in the appendix.

\begin{figure*}[t]
\begin{center}
\includegraphics[height=5.5cm]{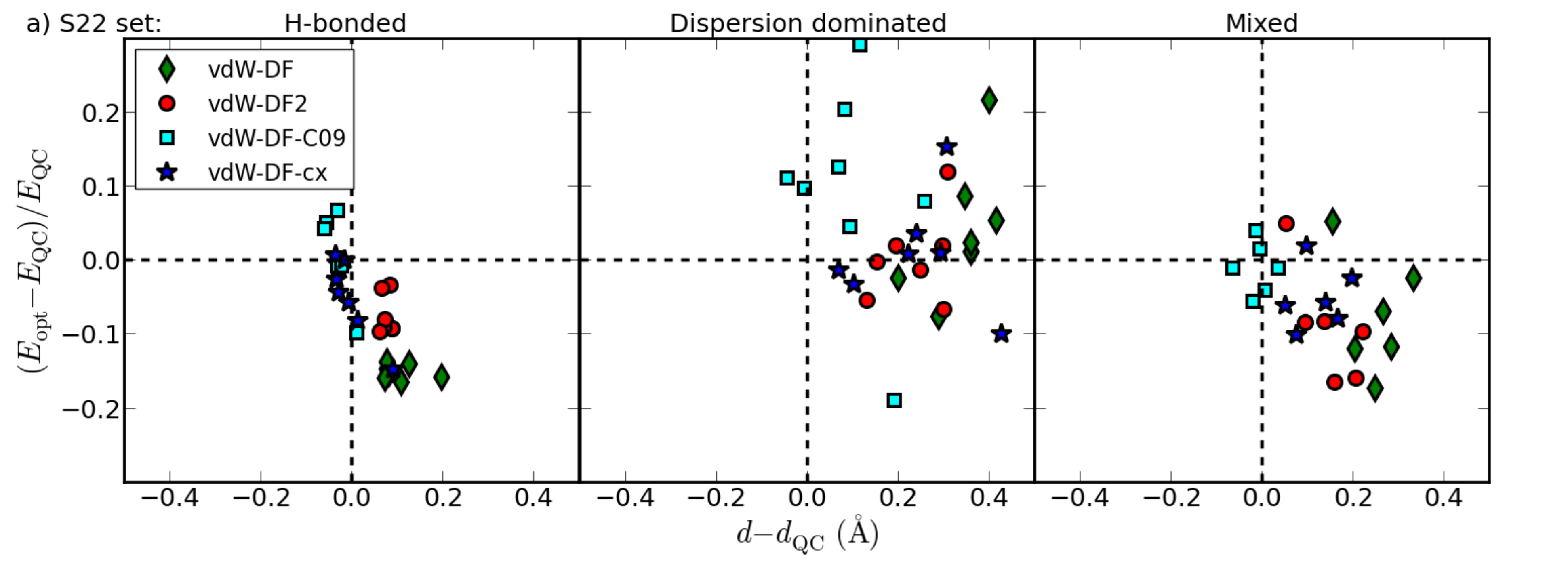}\\
\includegraphics[height=5.5cm]{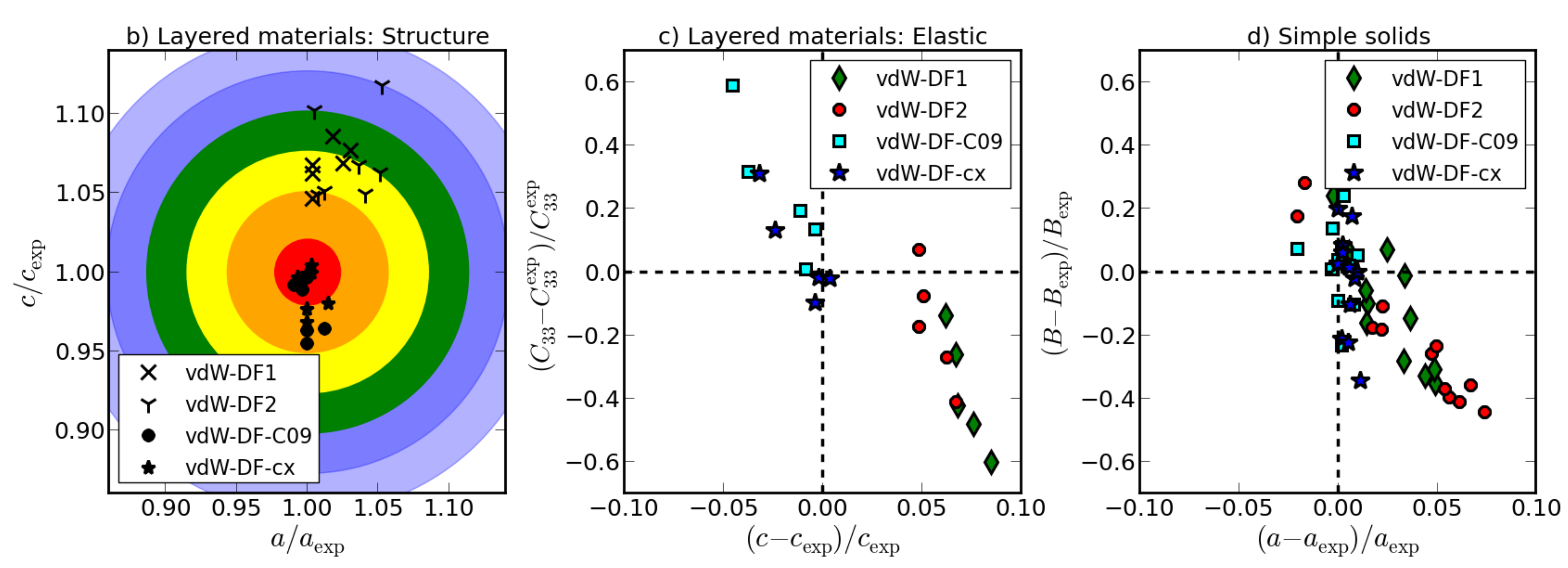}
\caption{\label{Fig1}
(Color online.) Scatter-plots summarizing the performance of vdW-DF versions\cite{Dion04p246401, Lee10p081101, Cooper10p161104, Berland13p} for a set of tests: (a) The S22 benchmark,\cite{Jurecka06p1985} where values for binding energy 
$E_{\rm opt}$ and separations $d$ of sets of molecular dimers are compared with values 
from accurate quantum-chemistry calculations; (b) Comparison of calculated and experimental 
values for in- and out-of-plane lattice constants $a,c$ for a group of layered (or intercalated) 
compounds; (c) Comparison of calculated and experimental values of the out-of-plane lattice constants 
and the out-of plane $C_{33}$ elastic modulus; and (d) Comparison of calculated and experimental 
values for the unit-cell lattice constant $a$ and bulk modulus $B$ for a group of simple solids (like Al, Ag, and Au). 
Most calculated results are from Ref.~\onlinecite{Berland13p}, but we have also included results for 
potassium intercalation in graphite and for the variation of Bi$_2$Te$_3$ layer binding energy with 
unit-cell size.}
\end{center}
\end{figure*} 
 
By \lq\lq{}electron gas tradition\rq\rq{} we mean that specific conservation laws are adhered to; including having an exchange-correlation (XC) hole that contains just one charge unit, and physical constraints built into the plasmon-pole response\cite{Dion04p246401,Thonhauser07p125112} that describes the nonlocal correlations. The recent most explicit functionals, vdW-DF1\cite{Dion04p246401, Thonhauser07p125112} and vdW-DF2,\cite{Lee10p081101} expand beyond the local electron-gas in terms of a plasmon propagator $S=1-1/\epsilon$ and arrive at a nonlocal correlation energy, $E^\nl_{\rm c}[n]$, that is quadratic in the density $n(\mathbf{r})$. The underlying plasmon-pole basis still reflects a collectivity that captures broader density variations.\cite{Dion04p246401,Rydberg01p,Berland13p205421, Hyldgaard13p} Seamless integration in the homogeneous electron gas (HEG) limit and the observance of physical constraints\cite{Gunnarsson76p4274,Langreth75p1425, Langreth77p2884,Dion04p246401,Hyldgaard13p, Gunnarsson79p3136} are features that vdW-DF shares with  LDA and GGA. 
The \textit{inner exchange\/} functional that describes the plasmon response giving rise to the nonlocal correlations in vdW-DF1 is also based on many-body theory in a diagrammatic form which was used in the design of early GGAs.\cite{Langreth87p497} 
vdW-DF2 is instead based on scaling laws for atoms and thus designed for higher accuracy for atoms and small molecules.
Two of us have recently designed an exchange functional LV-PW86r 
that extends the vdW-DF1 track.
The LV-PW86r exchange closely follows the vdW-DF1 inner functional 
up to moderate density variations, as described by the dimensionless parameter $s\propto|\nabla n|/n^{4/3}$, while switching to a better motivated and tested exchange description at large $s$ values.\cite{Berland13p} Paired with the vdW-DF1 correlation, we obtain an improved consistency between the exchange and correlation description and therefore we label the full functional vdW-DF-cx, where cx emphasizes the aim of using a consistent exchange description.

In the following sections, we examine the efficacy 
of a set of nonempirical vdW-DF versions. 
Specifically, we examine the recently developed vdW-DF-cx, and its 
precursor C09x. We aim to demonstrate that 
these 
functionals are capable of not only describing 
molecular problems but also are general purpose materials theory tools. 
These nonlocal functionals  
have an approximative conservation of the exchange-correlation hole. 
The vdW-DF-cx goes furthest in enforcing consistency between exchange and correlation and leverages the 
conservation of the full exchange-correlation hole instead of, for example, seeking to conserve 
the exchange hole separately from the nonlocal correlation. Such conservation suggests good transferability.

\section{Towards a general-purpose density functional}


\subsection{Plasmon theory of the electron gas}

A general-purpose density functional should be able to describe both molecular interactions and the binding within bulk materials. 
This has been a long-standing objective
for the development of nonlocal functionals by us\cite{Dion04p246401, Thonhauser07p125112, Langreth09p084203,Andersson96p102, Hult96p2029, Hult99p4708, Rydberg00p6997, Rydberg01p, Rydberg03p126402, Dion04p246401, Dion04p, Langreth05p599, Thonhauser07p125112, Langreth09p084203,Ziambaras07p155425} and others\cite{ Lazic12p424215}.
While vdW forces are generally expected to be important for molecular systems, nonlocal correlations are proving to be important for many other kinds of systems. 
For example, nonlocal correlations even 
play a significant role for the cohesion of covalently bound solids.\cite{Klimes11p195131,Berland13p}

The vdW-DF method aims 
to follow the same electron-gas traditions as 
the LDA and the GGA, by 
building on a GGA-type
plasmon-pole description.\cite{Rydberg00p6997,Rydberg01p,Dion04p246401,Thonhauser07p125112,Hyldgaard13p} 
Some details are presented in the appendix.

The plasmon response is described by an inner functional that reflects LDA and gradient-corrected exchange.\cite{Dion04p246401}
The nonlocal correlation can in turn be viewed as a counting of the plasmon zero-point energy shifts in the picture of Rapcewicz and Ashcroft.\cite{Rapcewicz91p4032}
The total energy also includes
an outer semi-local functional
defined as LDA plus gradient-corrected exchange.
This outer functional can also be represented in terms of  
having a plasmon response.
The vdW-DF versions, in general, all have deviations in the description of 
the exchange of the
inner and outer functionals, i.e.\ they have a nonzero value for
the cross-over term specified in \eq{eq:deltaExdef}.  Nevertheless, we
seek to achieve a consistent treatment of exchange,
that is, to use the same plasmon description for all functional components. 
Such consistency leads to an automatic conservation of the total exchange-correlation hole. 
This goal guided our design of vdW-DF-xc.

A fully consistent vdW-DF version, i.e.\ one with a good plasmon model, should be applicable 
for both
dense and sparse matter, and thus have the potential for extending the dramatic
success
of constraint-based DFT.\cite{Burke12p150901} 
The new version, vdW-DF-cx, goes furthest in 
enforcing
consistency between the inner and outer semilocal
functionals.\cite{Berland13p}
Given the emphasis on exchange consistency, the merits of a vdW-DF
version boils
down to exploring the quality of its plasmon-response
description.\cite{Berland13p}

\subsection{``Are we there yet?''}

Figure \ref{Fig1} presents an overview of the performance of different 
vdW-DF functionals based on (a) molecular dimers, (b,c) layered 
materials,\footnote{To the layered-system results presented in 
Ref.\ \protect\onlinecite{Berland13p}
we have added new results for C$_8$K (also reported in Table \protect\ref{tab:carbon}) and
for Bi$_2$Te$_3$.
The latter was obtained for normconserving pseudopotentials with full
relaxation
of both forces and stress. Our self-consistent
vdW-DF1/vdW-DF2/vdW-DF-C09/vdW-DF-cx
calculations determine in-plane lattice constants for Bi$_2$Te$_3$
$a=4.47/4.55/4.35/4.36$ {\AA},
out-of-plane lattice constant $c=11.03/10.85/10.08/10.13$, and elastic
constants
$C_{33}=19/28/48/43$ GPa. Corresponding experimental lattice constants
are $a_0=4.39$ {\AA},  $c_0=10.166$ {\AA}, 
Ref.\ \protect\onlinecite{Nakajima63p479}.
An experimental elastic constant $C_{33}^0=47.7$ GPa is listed in the
comparison
Ref.\ \protect\onlinecite{Bjorkman12p424218}.}
and (d) covalently-bound solids.  The overview suggests that 
the goal of designing
a general purpose functional that works simultaneously for dense and sparse matter is a realistic one. 
This is true even if the versions differ in accuracy 
since 
the newer versions like vdW-DF2 and vdW-DF-cx 
perform better than older functional like vdW-DF1 and earlier special-purpose functionals.
Nevertheless, the aim of a general-purpose functional should go beyond 
merely being able to describe different kinds of materials, like 
covalently-bonded solids and dispersion-bound molecular dimers, it should 
also be able to describe the delicate balance between interactions in 
systems where different forces or ground-state configurations compete. 

\begin{figure}[]
\includegraphics[width=0.96\columnwidth]{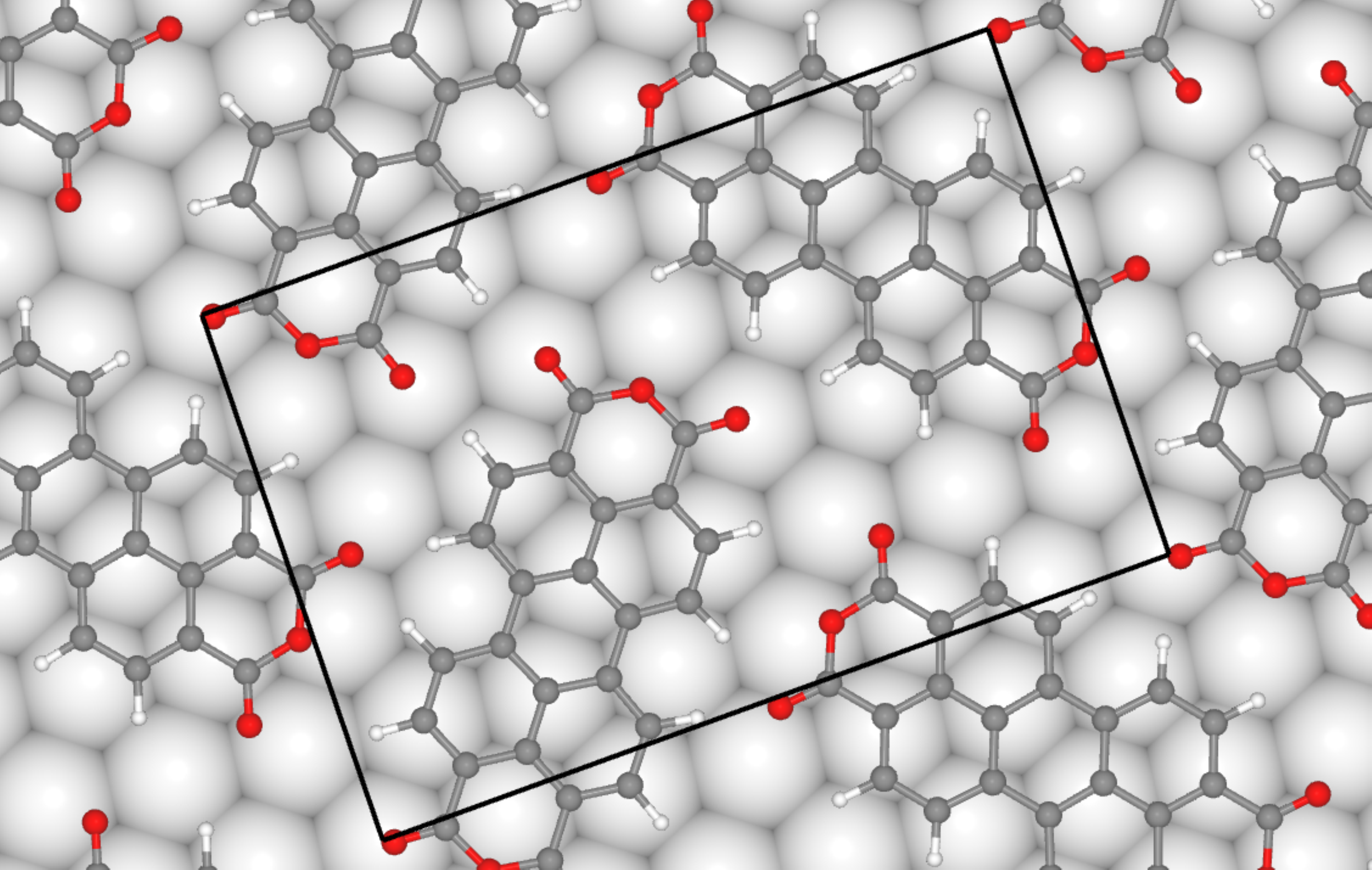}
\caption{
\label{Fig2}
(Color online.) The long-range ordered commensurate monolayer of PTCDA 
(perylene-3,4,9,10-tetracarboxylic-3,4,9,10-dianhydride) on Ag(111) 
taken from Ref.~\protect\onlinecite{Kraft2006}.  
Oxygen, carbon, hydrogen and silver atoms are represented by red, grey, 
white and light grey spheres, respectively.  In the unit cell 
($a =$ 19.0 \AA, $b$ = 12.6 \AA, $\gamma$ = 89$^o$), 
there are two flat lying PTCDA molecules in a herringbone arrangement.}
\end{figure}

We have identified and addressed a number of systems where vdW forces compete with other interactions; 
with interactions spanning from strong 
ionic bonds to covalently bonded solids and physisorbed molecules within pores and at surfaces. We explore 
basic physical properties of these materials, such as lattice constants and adsorption energies, 
as well as the competition between the paraelectric and ferroelectric phases in 
PbTiO$_3$ 
and the 
graphite and diamond phases of carbon. 
By examining these systems 
we show that 
vdW-DF-cx is indeed a general purpose tool and as such holds significant promise as 
a unified functional for exploring sparse and dense matter alike.

\section{Computational details}

For all systems, self-consistent vdW-DF calculations were performed using 
ultra-soft pseudopotentials (unless otherwise noted) as implemented in the 
\textsc{Quantum Espresso} DFT package.
 We have implemented the vdW-DF-cx functional in 
\textsc{Quantum Espresso}, which is now publicly available.
The energy cutoff and k-point mesh 
were chosen to converge the binding energy to within 1 meV. 
All atoms were relaxed until the Hellmann-Feynman 
forces were less than 15 meV/{\AA}.
We have used full stress relaxation\cite{Sabatini12p424209} of the unit cell unless otherwise noted.

For PbTiO$_3$ a five-atom unit cell was employed.

For the metal-organic-framework, we adopted the description in Ref.\ \onlinecite{Yao12p064302}, using 
a limited wavevector sampling, as motivated by the large hexagonal unit cell ($a=25.881$ {\AA} and 
$c=6.8789$ {\AA} for Mg-MOF74; $a=25.887$ {\AA} and $c=6.816$ {\AA} for Zn-MOF74).
These parameters reflect the experimental unit cell.  

The study of the adsorption of an organic molecule proceeded with a three-layer Ag(111) slab 
and 33 atoms per layer. The adsorption geometry is illustrated in the top panel of Fig.\ \ref{Fig2}. 



\section{Results: Testing nonlocal functionals when forces compete}

\subsection{A bulk-matter challenge: the ferroelectric response in PbTiO$_3$} 

\begin{figure}[]
\centering
\includegraphics[width=1.0\columnwidth]{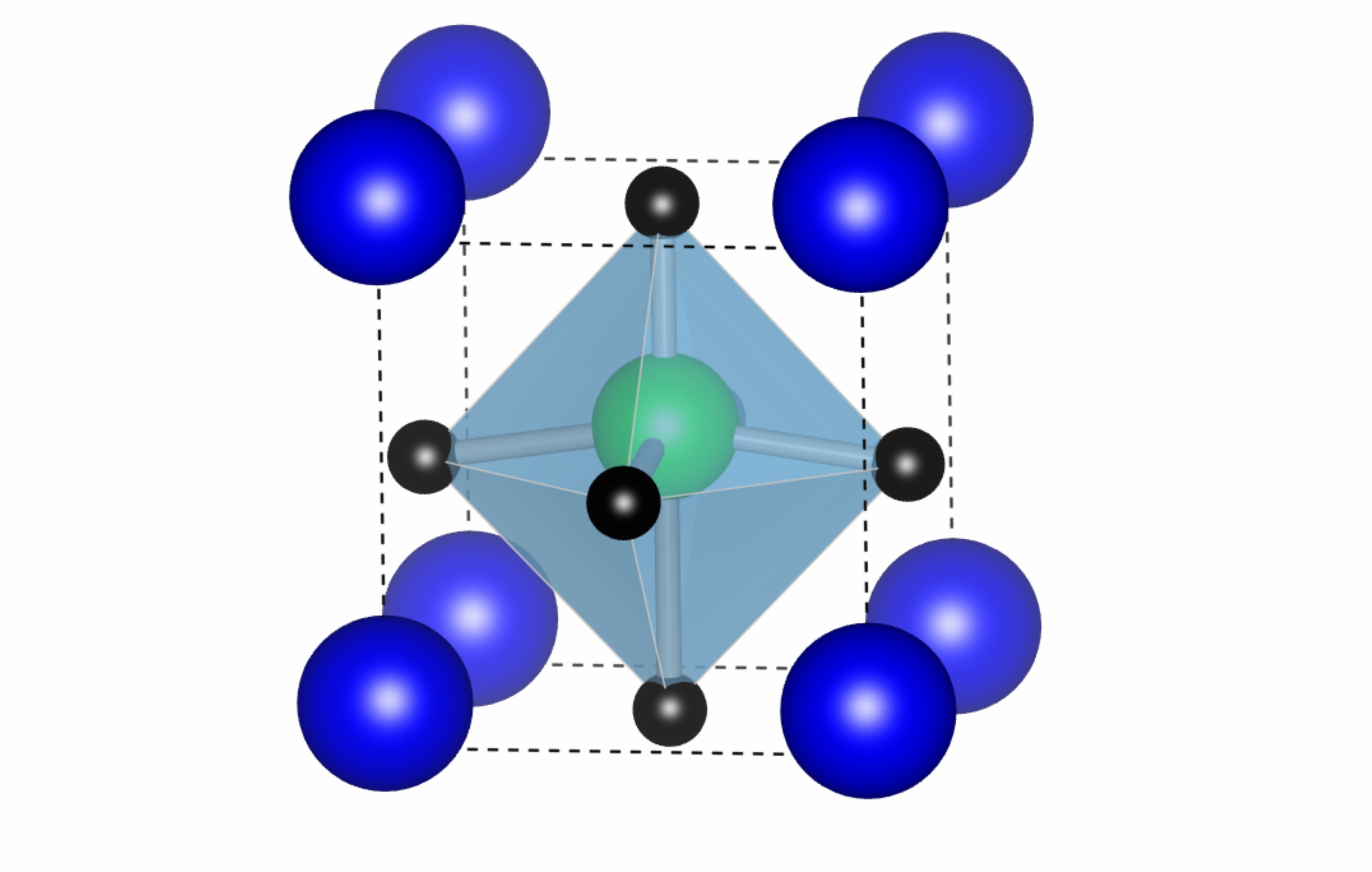}\\[1em]
\includegraphics[width=0.9\columnwidth]{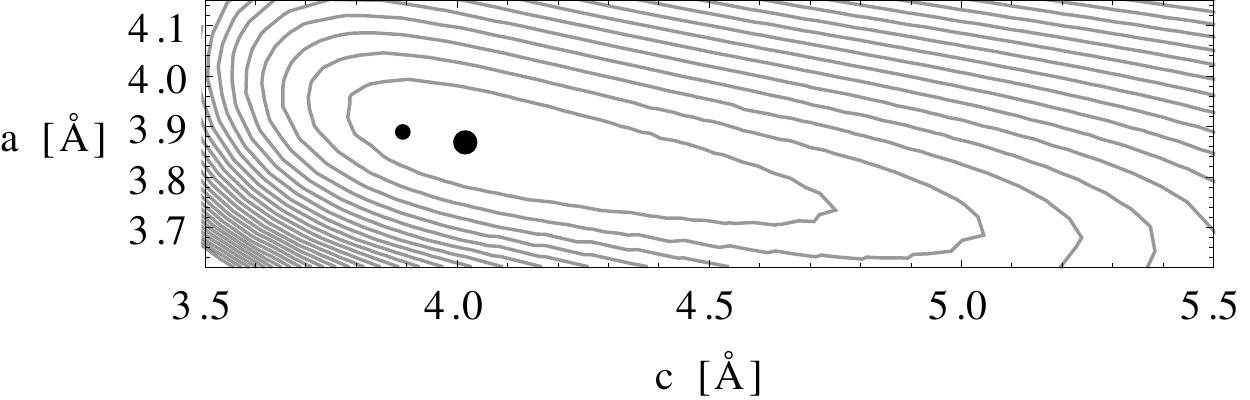}\\[1em]
\includegraphics[width=0.9\columnwidth]{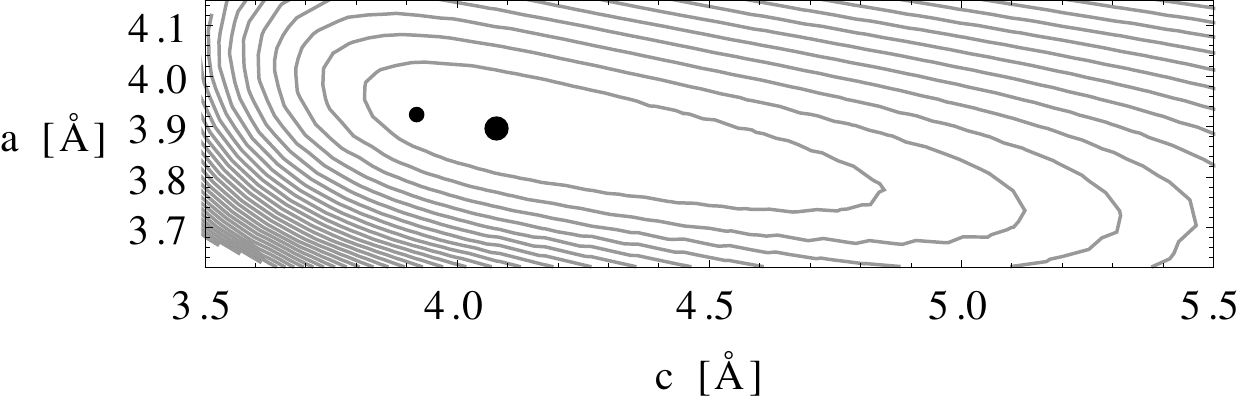}
\caption{
(Color online.) Unit-cell atomic configuration (top panel) and potential energy surface (PES) 
of PbTiO$_3$ (PTO) as evaluated in LDA (middle panel) and in vdW-DF-cx (lower 
panel).  The internal energy (that is, the DFT total energy) variations are mapped by
contours for the variation with unit-cell parameters `$c$' and `$a$'. 
The large (small) dot identifies the optimal geometry for the
tetragonal (simple-cubic) phase.  
The contours shown are separated by 0.1 eV relative to the DFT energy
of the tetragonal state. Both LDA and vdW-DF-cx predict the experimentally
correct polarization and both seem to have a good balance between exchange and
correlation.}
\label{fig:PTO}
\end{figure}
\begin{table*}
\caption{\label{tab:pto}Structural and electronic properties of PTO. $a$ and $c$ denote the equilibrium in-plane and out-of-plane lattice parameters. P$_z$,  $\delta E_{\rm phase}$ and $\delta E_{\rm phase}^{\rm c,nl}$ are the magnitude of the spontaneous polarization, the energy difference between the cubic and tetragonal phases and the non-local correlation compononent of the total energy, respectively.
The polarization vector $P$ is
aligned with the $c$ axis.  The table also lists atomic positions $\xi_z$
(the crystal-coordinate vectors projected on the polarization direction) for the tetragonal phase,
with the unit-cell offset as defined in Ref.\ \protect\onlinecite{Kuroiwa01p217601}.}
\begin{tabular*}{\textwidth}{@{}l@{\extracolsep{\fill}}dddddd@{}}\hline\hline

   & \multicolumn{1}{c}{vdW-DF1}

   & \multicolumn{1}{c}{vdW-DF2}

   & \multicolumn{1}{c}{vdW-DF-C09}

   & \multicolumn{1}{c}{vdW-DF-cx}

   & \multicolumn{1}{c}{Exp.}

   & \multicolumn{1}{c}{LDA}

\\\hline
$a$ [{\AA}]
& 3.8741 & 3.8893 & 3.8927 & 3.8955 & 3.9040^a
& 3.8663
\\
$c$ [{\AA}]
& 4.8152 & 4.8681 & 4.0518 & 4.0788 & 4.1575^a
& 4.0190
\\
P$_z$ [C/m$^2$]
& 1.22 & 1.19 & 0.72 & 0.75 & 0.75^b
& 0.74
\\
$\delta E_{\rm phase}$ [eV]
& -0.002 & -0.265 & -0.047 & -0.055 & -
& -0.053
\\
$\delta E_{\rm phase}^{\rm c,nl}$ [eV]
& 0.018 & 0.298 & 0.080 & 0.092 & -
&
\\
$\xi_z$(Pb)
& 0.16613  & 0.16328  &  0.09199  &  0.09676 &  0.1206(6)^a
&  0.09709
\\
$\xi_z$(Ti)
& 0.60949  & 0.60919  &  0.56489  &  0.56808 &  0.5773(4)^a
&  0.56592
\\
$\xi_z$(O$_{I}$)
& -0.02737 & -0.02864 & 0.00958 & 0.00853 & -0.0077(9)^a
&  0.01265
\\
$\xi_z$(O$_{II/III}$)
& 0.5 & 0.5 & 0.5 &  0.5 & 0.5^a
& 0.5
\\
\hline
\hline
\multicolumn{7}{l}{$^a$ Ref.\ \onlinecite{Kuroiwa01p217601}; \quad $^b$ Ref.\ \onlinecite{Gavrilyachenko70p1203}; \quad $^c$ Ref.\ \onlinecite{Li96p1433}.
$^d$ Ref.\ \onlinecite{King-Smith94p5828}; \quad $^e$ Ref.\ \onlinecite{Waghmare97p6161}.}
\end{tabular*}
\end{table*} 


$AB$O$_3$ perovskites oxides (see Fig.~\ref{fig:PTO}, top panel) are exciting materials.
A tremendous variety in their physical properties
can be achieved as a result of the choice of $A$ or $B$ site cations.
They can be metals, band insulators, Mott insulators, superconductors, 
magnetic and ferroelectric with tunable coupling between these degrees 
of freedom.\cite{Hwang12p103, Kalinin13p858}
Ferroelectrics, and by extension piezoelectrics, have been of particular 
interest due to their current and potential technological importance.  
Ferroelectrics are characterized by a spontaneous polarization which can 
be switched by an electric field, while piezoelectrics are a special 
subclass of ferroelectrics that exhibit a mechanical response that 
accompanies the change in polarization.

PbTiO$_3$ (PTO) has been extensively studied both by theory and experiment due to its importance as an end member of many high-response piezoelectrics. PTO has a single phase transition from a cubic, paraelectric phase (the ideal perovskite structure) to a tetragonal, ferroelectric phase (with a large $c$/$a$ ratio - see mid and lower panels of Fig.~\ref{fig:PTO}) at a relatively high transition temperature, $T_{\rm c} \approx 700$ K. 
PTO is a displacive ferroelectric in which the polarization is derived from the off-center displacements of the $A$ and $B$-site cations relative to their respective oxygen cages. The relatively large polarization of PTO is related to both the presence of a ferroelectrically active $A$-site cation and the partially covalent bonding of the Ti ions, which results in  anomalously large Born effective charges.\cite{Zhong94p3618}

Typically, DFT calculations of bulk oxide ferroelectrics have employed LDA for exchange and correlation. This is largely due to the fact that GGAs overestimation of the lattice constant results in it severely failing to describe the local structure of ferroelectric oxides.  LDA only modestly underestimates lattice constants and thus performs reasonably well.\cite{Wu05p37601,Wu04p104112} In any event, this description of the structure and energetics of the relevant phase is particularly problematic when attempting to predict $T_{\rm c}$ as they are closely related to both the polarization and the energy difference between the paraelectric and ferroelectric phases. Naturally, LDA consistently underestimates $T_{\rm c}$. Negative pressure simulations,\cite{Zhong94p1861, Zhong95p6301, Waghmare97p6161} and new functionals\cite{Wu06p235116} have been invoked to correct for these errors in DFT.

To describe PbTiO$_3$ accurately we must be able to both describe the 
ground-state structure and the atomic displacements as the material
makes the transition from the
ground-state ferroelectric phase to the high-temperature paraelectric phase.
Even though bulk-oxide ferroelectric forces are not typical sparse matter systems, 
we deem that they are good test systems for vdW-DF-cx. Due to strong competition 
between long range coulombic interactions and local covalent interactions, 
small differences in the nonlocal correlations energy could possibly tip the 
balance between the high-temperature cubic phase and the groundstate tetragonal phase.

Table \ref{tab:pto} details our comparison of the performance
of different functionals in terms of structure, thermal stability,
and the internal displacements of atoms.   
vdW-DF1 and vdW-DF2 significantly overestimate the $c$ lattice constant and predict dramatic internal distortions.
vdW-DF2 also predicts a very large energy difference between the phases.
vdW-DF-cx and vdW-DF1-C09 have the best performance in terms of displacements
and energetic difference of the phases.

The ground state structure has characteristic atom separations
of 2.25 {\AA} indicating that
nonlocal correlations can be important. Indeed, we find that for all considered vdW-DF versions the
energy difference between the $E_{c}^{\nl}$ contributions to the energy of the two phases is larger than the total change in internal energy.

\subsection{Small-molecular absorption: weak chemisorption in a metal-organic framework}

With the development of vdW-DF and other sparse-mater methods, DFT has 
become a key tool for the screening of potential MOF candidates 
for technologically-important applications such as hydrogen storage and 
gas sequestration for carbon capture.\cite{Furukawa13p974,Tranchemontagne12p13143,Zhou12p673}
Describing the adsorption within MOFs is however challenging because of the presence of metal ions which introduce an electrostatic component to the binding as well as
covalent effects. Due to these competing interactions, MOFs present an ideal testing ground for methods aiming to describe the cross over from weak, to moderate, and strong chemisorption.

\begin{figure}[]
\includegraphics[width=1.1\columnwidth]{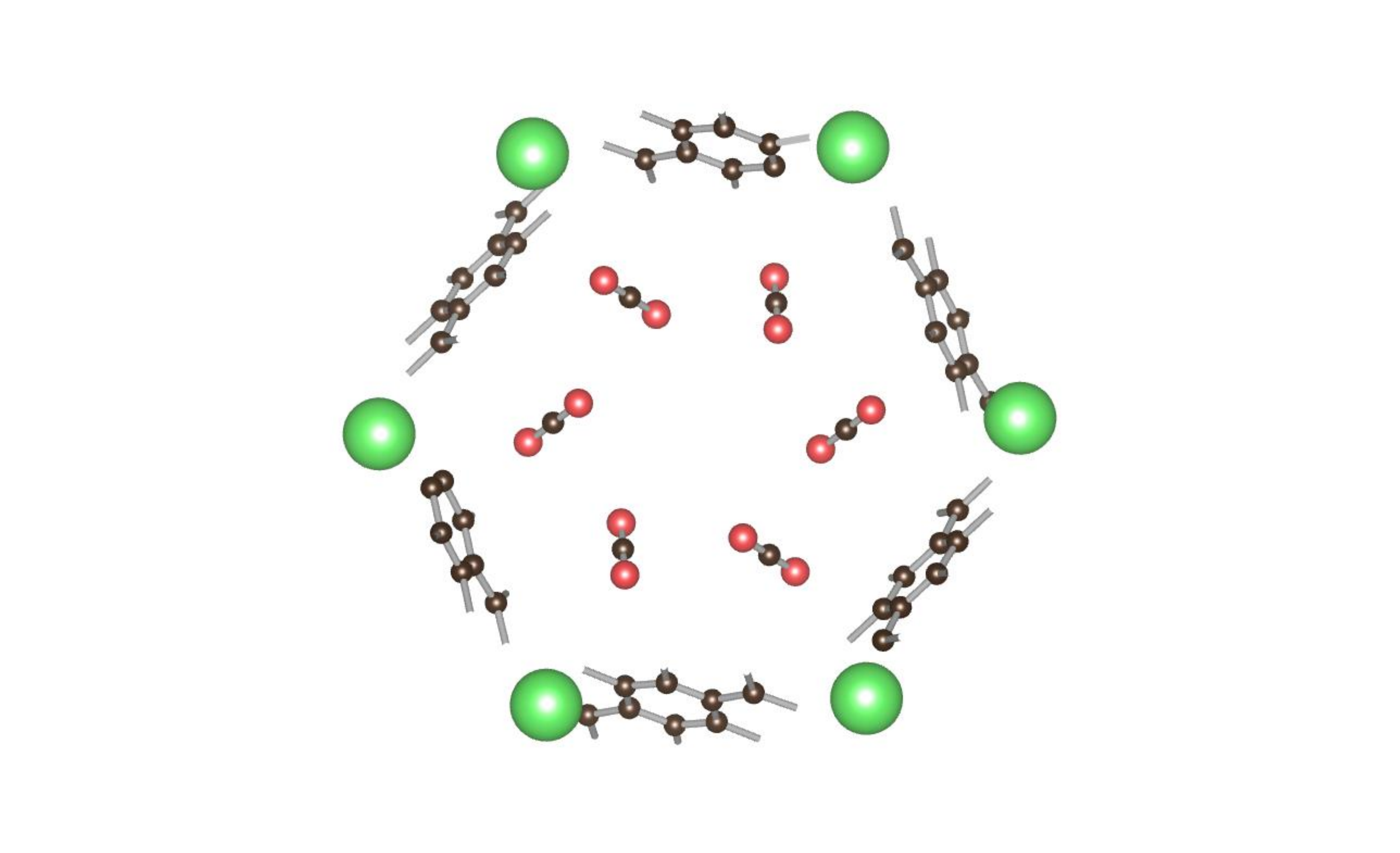}
\caption{
(Color online.) Final relaxed structure of the CO$_2$ adsorption in the Mg-MOF74 structure
as computed with vdW-DF-cx.  For all vdW-DF versions,
the MOF74 unit cell was kept fixed at the experimental value while all atoms 
were allowed to relax.
For the H$_2$ incorporation we found no discernable difference between
adsorbing a single or the full load of six absorbates. In the case of 
CO$_2$, the absorption energy was found to be 20 meV deeper with 6 instead
of a single molecule at the Mg site. Table \protect\ref{tab:mof} summarizes the details of
our absorption study and compares with experimental studies and results of other
methods where available.}
\label{fig:MOF}
\end{figure}

Figure~\ref{fig:MOF} shows a calculated binding geometry of CO$_2$ in Mg-MOF74. Similar to previous studies,\cite{Yao12p064302, Canepa13p13597}
our calculations show that the CO$_2$ adsorption in Mg-MOF74 is governed by vdW forces. 
However, replacing Mg in the same framework with Al can lead to strong chemisorption, which can even significantly deform the adsorbed molecule.\cite{Canepa13p13597}
In this study, we limit ourselves to assessing how vdW-DF-cx performs for the adsorption of H$_2$ and CO$_2$ within Zn-MOF74 and Mg-MOF74. 

Table~\ref{tab:MOF74} shows a comparison of the performance of different functionals. vdW-DF-cx like vdW-DF1 predicts similar binding energies, 
which are significantly overestimated for H$_2$ adsorption, but in good agreement with experiments for CO$_2$ adsorption. vdW-DF-cx predicts slightly shorter separations than vdW-DF2 yielding a slightly better agreement with the experimental separation for CO$_2$ in Mg-MOF74, the only system where this quantity is experimentally available. Overall vdW-DF2 is the best method for the considered systems. 
Like in earlier MOF studies,\cite{Canepa13p26102} the ability of vdW-DF functionals to be applicable on both small-molecule and
expanded-lattice scales should be noted.

\begin{table*}
\caption{\label{tab:mof}Adsorption energies $E_{\text{ads}}$ (in eV) and distance $d$ (in \AA) from the closest atom of the adsorbed 
molecule to the metal site for H$_2$ and CO$_2$ adsorbed in Mg-MOF74 and Zn-MOF74. 
For comparison, references to results from LDA, GGA, and PBE-D studies are also included
where available.} 
\begin{tabular*}{\textwidth}{@{}l@{\extracolsep{\fill}}lddddddd@{}}\hline\hline
   &
   & \multicolumn{1}{c}{vdW-DF1}
   & \multicolumn{1}{c}{vdW-DF2}
   & \multicolumn{1}{c}{vdW-DF-cx}
   & \multicolumn{1}{c}{Exp.}
   & \multicolumn{1}{c}{LDA}
   & \multicolumn{1}{c}{GGA}
   & \multicolumn{1}{c}{PBE-D}\\\hline
\bf H$_2$ in Zn-MOF74  & $d$                     &  3.223 &  2.848 &  2.736 & -        &  -       &  2.83^a  & -\\
                       & $E_{\rm ads}$           & -0.135 & -0.119 & -0.137 & -0.091^a & -0.228^a & -0.046^a & -\\[3ex]
\bf H$_2$ in Mg-MOF74  & $d$                     &  2.729 &  2.526 &  2.535 & -        &  -       &  2.54^a  & -\\
                       & $E_{\rm ads}$           & -0.163 & -0.155 & -0.171 & -0.105^a & -0.257^a & -0.062^a & -\\[3ex]
\bf CO$_2$ in Zn-MOF74 & $d$                     &  3.007 &  2.835 &  2.752 & -        & -        & -        &   2.71^b\\
                       & $E_{\rm ads}$           & -0.381 & -0.313 & -0.371 & -        & -        & -        & -0.309^b\\[3ex]
\bf CO$_2$ in Mg-MOF74 & $d$                     &  2.401 &  2.341 &  2.325 &  2.283^c & -        & -        &   2.45^b\\
                       & $E_{\rm ads}$           & -0.520 & -0.473 & -0.520 & -0.487^b & -        & -        & -0.429^b\\\hline\hline
\end{tabular*}
\flushleft $^a$Values taken from Ref.~[\onlinecite{Zhou08p15268}].\\
$^b$Values taken from Ref.~[\onlinecite{Park12p826}].\\
$^c$Values taken from Ref.~[\onlinecite{Queen11p24915}].
\label{tab:MOF74}
\end{table*}

\subsection{Graphite-intercalated system}

Graphite intercalation is relevant for battery operation 
since the alkali uptake can be electrochemically controlled. 
The resulting staging of the graphite matrix produces a dramatic 
increase in the graphene sheet-to-sheet separation in concert with 
a charge transfer to the sheets. The charge transfer is 
known to buckle the graphene sheets.  

This industrial relevance motivated an early vdW-DF1 study on the 
formation energy and response of C$_8$K.\cite{Ziambaras07p155425}
However, despite the charge transfer in this system, vdW-DF1 still overestimated the out-of-plane lattice constant by 0.2~\AA.
It is therefore interesting to apply the new vdW-DF-cx functional to this problem.
Our new vdW-DF1 calculations also supersedes the old ones because the new calculations are self-consistent and allow for atomic relaxations, capturing the charge-induced sheet buckling.

Table~\ref{tab:carbon} presents our results for C$_8$K and graphite. Overall we find that the description
of the potassium intercalation is
improved
by the self-consistent relaxation. 
We also find that 
vdW-DF-cx provides 
the most accurate description of the C$_8$K  structure and
behavior among the vdW-DF versions investigated.
 

\begin{table*}[]
\caption{Structure and binding of graphite intercalation and graphite.
Unit-cell parameters (in-plane lattice constant $a$ and average
separation $d_{C-C}$ between carbon sheets) are given in {\AA}. 
The graphite interlayer binding energy $\Delta E_{\rm bind}^{\rm lay}$ is
given in meV per graphene-sheet atom.
The graphite AA-vs-AB stacking fault energy $\Delta E_{\rm SF}^{\rm AA}$ and the 
graphite-versus-diamond internal-energy difference $\Delta E_{\rm phase}^{\rm G/3C}$
are given in meV per graphite unit cell.  The corresponding phase transition 
pressure $p$ is estimated by simply dividing by the vdW-DF-cx estimate for 
the graphite-to-diamond volume change per atom, $\delta V_C \approx 3$ {\AA}$^3$.  
The C$_8$K formation energy $\Delta E_{\rm form}^{\rm K atom}$
is given in eV per C$_8$K formula unit and given relative to the internal energy
of graphite and of free potassium atoms. In the LDA calculation (but not for 
any vdW-DF studies) we include the effect of a small spin-polarization
energy ($\sim 26$ meV) for these potassium atoms.  As indicated by a pair 
of '$NA$' entries, the PBE does not give any meaningful account of graphite 
binding (it is nominally computed as 1.2 meV per graphene-sheet atom) and 
there consequently exists no PBE account of the C$_8$K formation energy either.
Because the enthalpy difference between phases is so small in vdW-DF-cx we
have also carried out a check using normconserving pseudopotentials, with 
results given in square brackets.
}

\begin{tabular*}{\textwidth}{@{}l@{\extracolsep{\fill}}ldddddd@{}}\hline\hline

   &

   & \multicolumn{1}{c}{vdW-DF1}

   & \multicolumn{1}{c}{vdW-DF2}

   & \multicolumn{1}{c}{vdW-DF-cx}

   & \multicolumn{1}{c}{Exper.}

   & \multicolumn{1}{c}{LDA}

   & \multicolumn{1}{c}{GGA}\\\hline

{\bf Graphite}	 
& $a$ 
& 2.473 
& 2.478 
& 2.466 
& 2.459^a
& 2.466 
& 2.467 
\\
& d$_{\rm C-C}$
& 3.581 
& 3.517 
& 3.275 
& 3.336^a
& 3.325 
& 4.063 
\\
& $\Delta E_{\rm bind}^{\rm lay}$
& 55 
& 53 
& 66 
& 52^e
& 25 
& NA
\\
& $\Delta E_{\rm SF}^{\rm AA}$ 
& 19 
& 24 
& 47 
& 
& 39 
& 1.3 
\\
& $\Delta E_{\rm phase}^{\rm G/3C}$  
& 790 
& 1129 
& 7\, [30] 
& 
& -100 
& NA 
\\
& $\Delta p_{\rm phase}^{\rm G/3C}$  
& 11 
& 15 
& 0.09\, [0.4] 
& 0.7^{b,c}/1.4^d
& -1.4 
& NA
\\[3ex]
{\bf C$_8$K} 
& $a$ 
& 2.494 
& 2.497 
& 2.487 
& 2.480^f
& 2.485 
& 2.490 
\\
& $d_{\rm C-C}$ 
& 5.44
& 5.43
& 5.22 
& 5.35^g 
& 5.17 
& 5.37 
\\
& $\Delta E_{\rm form}^{\rm K atom}$ 
& 1.00 
& 0.93 
& 1.29 
& 1.24^h 
& 1.67 
& NA 
\\
\hline
\hline
\multicolumn{7}{l}
{
$^a$ Ref.\ \onlinecite{Baskin55p544}; \quad 
$^b$ Ref.\ \onlinecite{Bundy61p383}; \quad
$^c$ Ref.\ \onlinecite{Khaliullin11p693}; \quad
$^d$ Ref.\ \onlinecite{Berman55p333}; \quad
$^e$ Ref.\ \onlinecite{Zacharia04p155406}; \quad
$^f$ Ref.\ \onlinecite{Nixon69p1732}; \quad
$^g$ Ref.\ \onlinecite{Dresselhaus81p139}; \quad
$^h$ Ref.\ \onlinecite{Aronson68p434}. 

}\\
\end{tabular*}
\label{tab:carbon}
\end{table*}
\subsection{The graphite/diamond phase transition}

Graphite is the ground state among carbon allotropes --- but  
only just so, as the binding energy difference to diamond is small. 
It  takes a large pressure to produce diamonds directly
from graphite but that is due to a large kinetic barrier. 
From analysis of the high 
temperature phase behavior and other thermal behaviors it is possible to extract an estimate of what would be the equilibrium phase difference in enthalpy or equivalently an estimate of what pressure is required to induce a phase transition (ignoring the kinetic barrier). 
In this procedure, the phase separation is estimated to have approximately a 
0.7 GPa equilibrium transformation pressure.\cite{Bundy61p383,Khaliullin11p693} 

Since graphite is a partly vdW bonded systems with sp$^2$ bonds between the atoms of the graphene sheet, while diamond is purely sp$^3$ bonded, it becomes interesting to test the
accuracy of a set of nonempirical
descriptions of this transition. 

Our results are summarized in Table~\ref{tab:carbon}. 
Interestingly, we find that while
LDA does provide a finite binding between the layers, it also  produces an incorrect
ordering of the enthalpy of graphite and diamond phases.
The set of nonlocal functionals all produce a binding of the graphite layers, although
vdW-DF1 and vdW-DF2 have a very high value for the predicted differences in the phase
enthalpy.

Overall we observe that vdW-DF-cx produces a prediction for the energy 
ordering that is closest to the carbon-allotrope behavior. 
However, we note we have not at this stage added lattice zero-point energy effects.

\subsection{Organic-molecule adsorption on Ag(111)}

The adsorption of organic molecules on coinage metal surfaces is a
problem where different sparse matter methods can yield widely varying
results, ranging from covalent binding with short molecule-to-surface
separations to a purely dispersion-bounded description as typically
predicted with vdW-DF1.~\cite{Berland09p155431, Mura10p4759,
  Li12p121409, Lee12p424213} These systems are therefore particularly
challenging ones.  

A prototypical model system is the
3,4,9,10-perylene-tetracarboxylic-dianhydride (PTCDA) molecule on
Ag(111), which forms a long-range ordered commensurate monolayer and
has been studied extensively both experimentally and
theoretically.~\cite{Glockler1998, Eremtchenko2003, Kraft2006,
  Sachs09p144701, Romaner09p053010, Bjork10p3407, Tkatchenko2010,
  Ruiz12p146103} 
  Results for the lateral adsorption structure (Figure~\ref{Fig2}) and vertical adsorption bond lengths are available. The system therefore represents a valuable benchmarking system.

 \begin{figure}[b]
\includegraphics[width=0.8\columnwidth]{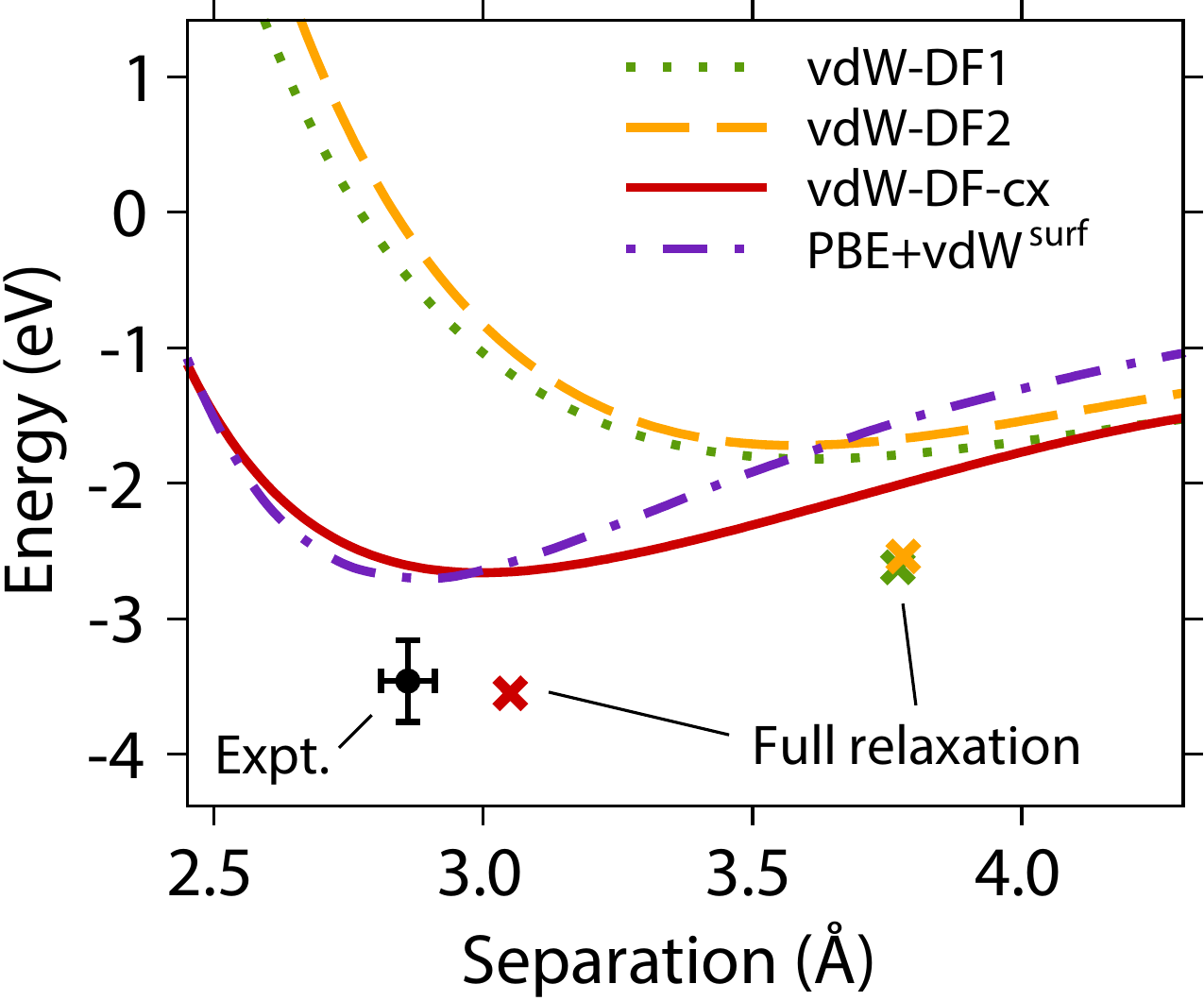}
\caption{\label{PTCDA} (Color online.) Potential energy curves (PECs)
  of the epitaxial monolayer of PTCDA on a Ag(111) substrate as a
  function of the vertical separation calculated by using vdW-DF
  versions and PBE+vdW$^\mathrm{surf}$, assuming flat and rigid molecules and surface.  The cross marks are
  adsorption energies obtained by performing full relaxation of the
  interface, both the molecule and the first atomic layer of the slab,
  and by including the intermolecular interaction properly.  The
  experimental values for binding energy and binding separation are
  shown as a black dot with error-bars.  The PECs are calculated in
  the same three-layer slab representation that was used in an earlier
  empirical pair-potential plus image-plane correction
  study.\protect\cite{Ruiz12p146103} }
\end{figure}
For a straightforward comparison with a previous
work~\cite{Ruiz12p146103} by Ruiz and co-workers, we first assume
that both the surface and the molecules are flat and rigid, before
doing a full geometry optimization.  The Ag substrate structure
is kept frozen at the atomic configuration obtained from experimental
lattice parameters of bulk Ag.  The potential energy curves (PESs) of
the PTCDA monolayer as a function of the vertical separation of the
layer are compared in Fig.~\ref{PTCDA}, along with the results from
Ref.~\onlinecite{Ruiz12p146103}.  The fully relaxed results, including
both the relaxation of the interface and the intermolecular
interaction energy between PTCDA molecules within the layer, are indicated by cross-marks in Fig.~\ref{PTCDA} and will be discussed further below.

The vdW-DF-cx performs similar to PBE+vdW$^\mathrm{surf}$.  Both show
good agreement with experimental binding separation, while vdW-DF1
and vdW-DF2 overestimate it.  A comparison of adsorption energy is,
however, not straightforward. Determining the adsorption energy of
PTCDA on Ag(111) is challenging and we only have a rough estimate for
an experimental value.  Standard temperature-programmed
desorption (TPD) is unapplicable because the molecule cracks before
desorption on heating.~\cite{Umbach1996} An estimate of 2.4 eV 
by Tkatchenko and co-workers~\cite{Tkatchenko2010} is given by 
two times of the desorption energy of a smaller but similar molecule
called NTCDA, which has about a half the size of
PTCDA.~\cite{Stahl1998}

We argue, however, that the binding energy may be significantly
larger for the following reasons: First, the desorption energy of
NTCDA (1.16 eV) was obtained~\cite{Stahl1998} for a loosely packed
monolayer at low coverage, in which intermolecular hydrogen bonds are
weak or missing.  According to our calculation using vdW-DF-cx, the
intermolecular interactions contribute to the desorption energy by
0.46 eV per molecule.  Further, Fichthorn and co-workers
showed~\cite{Fichthorn2002} that for large molecules
the pre-exponential factor
in Redhead formula for TPD should be several orders of magnitude
larger than a typical value for smaller ones due to a large
entropy associated with many local minima.  The use of a typical
pre-exponential factor underestimates the desorption energy of dodecan
(C$_{12}$H$_{26}$) on Au(111)\cite{Fichthorn2002} by 25\%.  Applying the
same ratio of underestimation and by including the intermolecular interaction energy
correction, we estimate the binding energy to be 3.46 eV (2.4 $\times$
1.25 + 0.46 eV).  The corresponding adsorption energy predicted by 
vdW-DF-cx falls within this experimental estimate, which presumably will be the case for PBE+vdW$^\mathrm{surf}$ as well. vdW-DF1 and vdW-DF2 predict smaller energies.

In order to quantify the degree of structural change of the molecule
and the surface separately, we first optimized the structure of the
molecule while keeping the flat surface intact by using
vdW-DF-cx. Then, a full relaxation of the interface has been performed
including the first atomic layer of the surface, while the
bottom two atomic layers are fixed.  

The energy gain by molecule relaxation, 0.36~eV per molecule, is much
larger than the gain by surface relaxation, 0.07~eV per molecule.
After full relaxation of the interface and including the
intermolecular interaction energy, the calculated binding energy is
3.55~eV. \footnote{ The average height of the anhydride oxygen atoms, carbon atoms, and
carboxyl oxygen atoms of the molecule are 3.09~\AA , 3.06~\AA, and 2.97~\AA,
respectively.  In comparison with experimental values of 2.97~\AA, 2.86
\AA, and 2.68~\AA, they are overestimated by 0.12~\AA, 0.20~\AA, and 0.29~\AA.
Including relaxations for the bare surface, the first atomic layer of the surface moved up out of the substrate by 0.020~\AA~on average.  After adsorption of the
molecule, however, the layer is pushed back down into the surface by 0.016~\AA\
on average.  The buckling of the surface, measured as the difference
of the minimum and the maximum vertical displacement of the atoms is
0.09~\AA~as compared to 0.18~\AA~for the molecule. }
We obtained similar amount of stabilization with vdW-DF1 and
vdW-DF2 as well.  


In this system, it is the nonlocal correlation and a weak
chemisorption component that pulls the molecules close to the Ag
substrate.  Semi-local functionals like PBE cannot accurately describe
this adsorption problem; the van der Waals forces are needed to pull
the molecule closer to the surface.

For some systems, we need a reliable account of the covalent bonds of
the surface to accurately describe the adsorption of
molecules.\cite{Klimes10p22201,Klimes11p195131} 
This is important for
accurately capturing the adsorption-induced surface relaxations, though this effect is less important for the adsorption case studied here.
The
vdW-DF-cx version can reliably describe both the molecular adhesion
and can, for example, accurately describe the Ag bulk
structure. For the asorption-system studied here, adsorption-induced relaxations of the surface are moderate.

\section{Discussion of vdW-DF versions} 

We here discuss the nature of the different nonempirical vdW-DFs developed in our collaboration. This serves to put vdW-DF-cx in the context of these earlier developments.

\subsection{vdW-DF1}

vdW-DF1\cite{Dion04p246401} uses revPBE\cite{Zhang98p890} as the exchange component of the outer
functional $E_{xc}^{\msl}[n]$ that describes most of the exchange part of the total energy. 
This choice was made because revPBE 
is a well tested functional that in practice ensures that no unphysical binding arises from exchange 
alone as, for example, in the case of the benzene dimer.\cite{Rydberg03p126402,Dion04p246401,Langreth05p599, Wu01p8748}

The exchange functional is usually described in terms of the reduced density gradient
$s  = |\nabla{n}|/2 k_{\rm F} n$ and the exchange enhancement factor {$F_{\rm x}(s)$}.
The enhancement factor of revPBE is similar to that of PBE, in particular they agree for
small $s$-values.  As an alternative to the constraint-based derivation, the PBE functional can
be justified by its similarity to the numerical GGA.\cite{Perdew96p16533}
This GGA form is based on imposing a cutoff of the gradient expansion approximation (GEA) for
the exchange-correlation hole to preserve its unit charge, a procedure that among other
things generalizes the analysis in the construction of the PW86 exchange.\cite{Perdew86p8800,Murray09p2754}
vdW-DF1 therefore benefits from some of the insights that underpin the construction of the PBE.
However, we stress that this design guide is implemented only for the exchange content in $E_{xc}^{\msl}$,
which ignores residual exchange in $E^{\nl}_c$. We also note that a modulation factor of the GEA hole
was used so that its exchange form differs from
a direct implementation of exchange-hole conservation\cite{Murray09p2754} at small $s$ values;
the large-$s$ enhancement factor of the numerical GGA was ignored altogether
in specifying the PBE exchange.

Nevertheless, vdW-DF1 successfully describes the binding in many dispersion bound 
(sparse matter) systems.\cite{Langreth09p084203,Berland09p155431} 
For instance, it does quite well for computing the energetics of layered systems,\cite{Kleis08p205422,Berland13p,Bjorkman12p424218} as illustrated by the graphite case in Tab.~\ref{tab:carbon}.
The fact that revPBE is quite repulsive, however, causes
vdW-DF1 to systematically produce too large separations 
in molecular systems 
and layered
materials.\cite{Murray09p2754,Berland11p1800,Klimes11p195131,Berland13p} 
This is illustrated respectively by the S22 data shown in Fig.~1 a) and the for layered systems in b).
In addition, the lattice constants of many inorganic solids are overestimated, as seen 
Fig.~\ref{Fig1}~d).  The latter can be an issue when modelling the adsorption of molecules, 
because it can be important to permit 
the near-surface atoms to relax.\cite{Klimes11p195131} 
The PBE\cite{Perdew96p3865} functional also tends to overestimate 
lattice constants, a feature that has been linked to the fact that its enhancement factor differs from 
that obtained from an analysis of the many-body diagrams for the weakly perturbed 
HEG.\cite{Rasolt75p1234,Langreth81p446,Langreth87p497,Staroverov04p75102}
It is plausible that the lattice-constant overestimation is exacerbated in vdW-DF1 because
it does not retain a LDA-type description of linear response and has a different
balance between gradient corrections to exchange and 
correlation.\cite{Staroverov04p75102}

\subsection{vdW-DF2} 

vdW-DF2 employs an exchange-enhancement description for the inner functional $E_{xc}^{0,i}[n]$ 
obtained from the formal results of Schwinger and of Elliott and 
Burke.\cite{Schwinger80p1827,Elliott09p1485} This is
in contrast to using the many-body results of the HEG to describe the plasmon response, as was done
in the design of the vdW-DF1. The approach is demonstrated to be accurate for 
atoms\cite{Perdew06p223002} and can be shown\cite{Elliott09p1485} 
to essentially have the character of the Becke-88 functional\cite{Becke88p3098} for small values of the 
scaled gradient. This is one factor indicating good performance for molecules. 

vdW-DF2 also updates the choice of exchange component of the outer functional 
to the revised PW86.\cite{Murray09p2754}
The enhancement factor of this functional arises from enforcing a hard cutoff on the exchange-hole 
of GEA.\cite{Perdew86p8800} Since it has been shown that this exchange choice agrees well with that of 
a Hartree-Fock description of exchange effects between molecules,\cite{Murray09p2754, Kannemann09p719} 
one can expect vdW-DF2 to have high accuracy and good transferability for
small-molecular systems.\cite{Lee10p081101}

vdW-DF2 indeed performs well for many types of systems, including dimers\cite{Lee10p081101}
and the adsorption of small molecules.\cite{Berland11p1800,Lee11p193408,Lee12p424213} It significantly
improves upon the account of vdW-DF1. 
These trends are captured in the scatter plots for the S22 data set in Fig.~\ref{Fig1}~a), 
as well as for the adsorption in MOF74 listed in tab~\ref{tab:MOF74}.
However, vdW-DF2 only slightly improves interlayer separations of
layered compounds relative to vdW-DF1,\cite{Bjorkman12p424218} as shown in 
Fig.~\ref{Fig1} b). Further,
the adsorption energy for bigger molecules can be somewhat underestimated,\cite{Berland13p205421} 
as seen in the Fig.~\ref{PTCDA} 
and the value of lattice constants of solids usually worsens, 
Fig.~\ref{Fig1}~d).\cite{Klimes11p195131, Berland13p}
This lack of improvement for solids and larger molecular systems is expected since neither 
the inner nor the outer functional have
an exchange-enhancement factor that is consistent with the results of a many-body physics analysis for a weakly
perturbed electron gas.\cite{Rasolt75p1234,Langreth81p446,Langreth87p497,Langreth90p175,Staroverov04p75102}

We note that vdW-DF2 has fair consistency between the inner exchange parameterizing the plasmons
(the plasmons that set the determination of nonlocal correlation $E_c^{\nl}$) and the
exchange part of the outer functional $E_{xc}^{\msl}$ (the semilocal part of the
full exchange-correlation energy) at least at small to moderate values of the reduced gradient $s$.
This fair consistency was not used to motivate the vdW-DF2 design but follows from the fact that it
systematically implements GGA-type descriptions that are excellent for (small-)molecule-type problems.

\subsection{vdW-DF-cx}

vdW-DF-cx\cite{Berland13p} effectively implements 
the conditions necessary for consistency between the inner and outer exchange. Starting with the 
vdW-DF1 plasmon-pole description of the inner-functional and hence nonlocal correlations, 
Ref.\ \onlinecite{Berland13p} demonstrates that the range of $s$-values that contributes to 
$\Enl$ is limited to $s< 2-3$ for most material properties of interest.  
It is therefore sufficient to demand consistency of the inner and outer functional descriptions in 
this limited regime. At the same time, the $E^{\nl}_c$ analysis of the relevant contributing $s$ 
values also implies that for $s > 2-3$ one can proceed with a traditional (numerical-GGA) 
analysis,\cite{Perdew86p8800,Murray09p2754} to specify an exchange enhancement factor for the large-$s$ 
regime. Overall we arrive at an outer-functional exchange specification, termed LV-rPW86, that 
is designed exclusively with the purpose of working with the vdW-DF1 description of nonlocal 
correlations. The result is a nonempirical vdW-DF version, vdW-DF-cx\cite{Berland13p} 
that effectively ensures hole conservation for most materials problems; the only exception 
being vdW binding of noble gas atoms and small molecules, for reasons discussed in Ref.\ [\onlinecite{Berland13p}].

Since the plasmons giving rise to the nonlocal correlation of vdW-DF1 are described by a
near-HEG behavior,\cite{Berland13p} vdW-DF-cx has an exchange component which is not far from that of
previous exchange functionals used with vdW-DF1, like C09x\cite{Cooper10p161104} and optB86b.\cite{Klimes11p195131}
Both of these functionals resemble the design logic of the PBEsol functional\cite{Staroverov04p75102} for 
small-to-medium $s$ values (but with different large-$s$ tails) and have significant improvements in 
lattice-constants over vdW-DF1.  Albeit, neither C09x nor optB86b were designed to minimize the 
value of the cross-over term expressed in \eq{eq:deltaExdef}, and thus cannot leverage the conservation 
of the total exchange-correlation hole that is implied (for a plasmon-pole description) 
in \eq{eq:effectiveCriterial}.

Given the strong connection with PBEsol it is not surprising 
that the new vdW-DF-cx functional improves upon lattice constants for both bulk and layered 
vdW systems, as shown in the lower panels of Fig~\ref{Fig1}.\cite{Berland13p} 
For systems involving the smallest of 
molecules, like H$_2$ within MOF74, vdW-DF2 is more appropriate.
Nevertheless, vdW-DF-cx comes out in overall slightly better than vdW-DF2 for the S22
benchmark set of molecular dimers.
Systems characterized by competing interactions are often those where vdW-DF-cx does particularly well.  
In addition to the importance of using a consistent exchange account, this trend can be related to that these systems are characterized by shorter separations that the purely vdW bonded ones. These shorter separations make the near-HEG description that both the correlation and exchange of vdW-DF-cx rely on more appropriate.  
It is encouraging that vdW-DF-cx improves 
the description of binding energies in systems that range from bulk systems, over layered 
compounds and to molecular systems, as well as to systems characterized by competing interactions. 


\section{Conclusion}

Good tools are valuable for the theoretical description and exploration of
general matter; i.e.\ systems that are comprised of regions of both dense and sparse electron
concentrations. For ground-state properties DFT has been available for almost half a
century. During the last decade its application to sparse (and hence general) matter
has been significantly improved. The variety of relevant systems is enormous, far beyond
what is indicated by the applications in this paper: the ferroelectric response in PbTiO$_3$,
the adsorption of small molecules in metal-organic frameworks (MOFs), the phase transition between
graphite and diamond, and the adsorption of an organic-molecule on the Ag(111) surface.  Nevertheless, the
here selected examples are sufficient enough to represent the variety of possible materials challenges.
 
The nonempirical vdW-DF method, the tool of this study, is characterized by high
ambitions of simplicity and  physicality. The applications are performed
with four vdW-DF functionals, vdW-DF1, vdW-DF2, vdW-DF using C09 exchange,
and the very recent vdW-DF-cx.\cite{Berland13p}  In a revitalization of the
development of the vdW-DF method we have particularly noticed that it benefits from extensive
studies of the almost homogeneous electron gas, including those leading up to the
GGAs. This is the electron-gas tradition of a plasmon-pole 
description.\cite{Lundqvist68p117, Lundqvist67p193, Lundqvist67p206,Gunnarsson76p4274,Langreth77p2884,Dion04p246401}
Our recent analysis\cite{Berland13p} shows that this plasmon description
should be used within vdW-DF to give a good description of \emph{both} exchange
and correlation effects.

The vdW-DF-cx functional has recently been developed on this ground and has been shown
to perform well for solids, layered materials, and for the S22 benchmark set.
In particular the good results for lattice parameters and elastic
response  should be stressed. Here, our results for vdW-DF-cx demonstrate that this functional is capable of accurately describing the structure and properties of a wide range of systems; ranging bulk oxides, to molecules adsorbed at surfaces and in porous media and for understanding the phase transition between a covalently bonded bulk solid to a dispersion bound layered material. 

Given the 
adherences to various conservation rules and the
associated potential for transferability, we believe that there 
are grounds for making some broader conclusions about the capability of the vdW-DF method. 
In fact, since the tests cases can be seen as difficult, covering a range of problems where 
interactions compete, there is a potential for fair performance also for general problems. 
At the very least, these results encourages us to test that conjecture in upcoming works. As such, in addition to providing a very competitive density functional for sparse matter, this work highlights the promise for further improving functionals, thanks to the robust and flexible formulation of the vdW-DF method.

\section{Acknowledgment} The authors thanks H.\ Rydberg and P.\ Erhart 
for useful discussions. Work by KB, ES, and PH  was
supported by the Swedish research council (VR) under grants
VR-2011-4052 and VR-2010-4149 and by the Chalmers Area of Advance, Materials.
Work by BIL was supported via the pension from the Swedish pension system. Work at Wake Forest University was entirely supported by
Department of Energy Grant No.\ DE-FG02-08ER46491. VRC was
supported by the Materials Sciences and Engineering Division,
Office of Basic Energy Sciences, U.S. Department of Energy. We
are also grateful for allocation of computational resources by
the Swedish National Infrastructure for Computing (SNIC) and by
Wake Forest University.

\appendix
 
\section{Formal theory}

\subsection{The vdW-DF framework}

The vdW-DF framework is rooted in the adiabatic-connection formula (ACF). The ACF embodies the fact that many-body interactions are reflected in the way a system responds
to changes in the potential. 

The ACF links the exchange-correlation energy to the reducible and irreducible 
density-density correlation functions $\chi_\lambda(\omega)$ and $\tilde{\chi}_{\lambda} (\omega)$, respectively, at a given coupling constant $\lambda$ 
 --- which respectively describe the induced charge given by an external and local potential ---
through an integral over the coupling-constant. 
Expressed as an integration over imaginary frequencies $u$,
\cite{Langreth75p1425, Gunnarsson76p4274, Langreth77p2884, Pines89p}
it reads 
\begin{equation}
E_{xc} + E_{\rm self}
= - \int_0^1 d\lambda \, \int_0^{\infty} \frac{du}{2\pi} \,
\text{Tr}\{ \chi_\lambda(iu) V \}\;.
\label{eq:ACFexpress}
\end{equation}
The infinite self-energy term $E_{\self}$ cancels out
 a corresponding
divergence in the right hand side of
\eq{eq:ACFexpress}.

In the vdW-DF framework, the ACF is recast\cite{Hyldgaard13p}
to include the coupling-constant implicitly within an effective longitudinal dielectric function $\kappa_{\ACF}(\im u)$, as follows
\begin{equation}
  E_{xc} + E_{\self}  =   \int_{0}^{\infty} \, \frac{du}{2\pi}
  \hbox{Tr} \{ \ln ( \kappa_{\ACF}(\im u)) \} \; .
  \label{eq:FirstExcRecast} 
\end{equation}
The effective dielectric function $\kappa_{\ACF}(\im u)$ is 
defined 
by a longitudinal projection of a scalar dielectric function  $\kappa_{\ACF}(\im u)=\nabla \epsilon(\im u) \cdot \nabla G \, $.
We also define an effective (coupling-constant averaged) local field response $\tilde{\chi}_{\rm ACF}$ using $\kappa_{\rm ACF}(iu)\equiv 1-V\tilde{\chi}_{\rm ACF}(iu)\ $.

In principle Eq.~(\ref{eq:FirstExcRecast}) can be made exact. \footnote{One can select a scalar, nonlocal dielectric constant $\epsilon(iu)$ that satisfies\cite{Rydberg01p}
\eq{eq:FirstExcRecast} for any given $E_{xc}$}
In practice, an approximate scheme for the scalar dielectric function $\epsilon$ is employed through 
a single plasmon-pole approximation for the plasmon propagator
$S(\omega) \equiv 1-\epsilon^{-1}(\omega)$. 
The form of $S$ is designed to observe all known conservation laws for the plasmons.\cite{Dion04p246401}

\subsection{Exchange-correlation hole conservation}

The relation
\begin{eqnarray}
E_{xc} & = & \frac{1}{2} \int d\mathbf{r}\, n(\mathbf{r}) \int \diff {\bf u} \, \frac{1}{4\pi u}n_{xc}(\mathbf{r}; {\bf u} ) \nonumber\\
& = & \int_0^{\infty} \frac{du}{2\pi} \, \hbox{Tr} \{ \ln(\kappa_{\ACF}(\im u)) \} - E_{\self}
\label{eq:varepsievalA}
\end{eqnarray}
links $\kappa_{\ACF}$ to $E_{xc}({\bf r})$
and hence to an integral over the exchange-correlation hole $n_{xc}({\bf r},{\bf u})$.

The exchange-correlation hole is conserved if 
\begin{align}
\int \, d\mathbf{u} \, n_{xc}(\mathbf{r}; \mathbf{u}) 
\equiv n_{xc}(\mathbf{r}; \mathbf{q}'=0) =  -1\,.
\label{eq:conserveXChole}
\end{align}
This  condition 
can be formulated
\begin{eqnarray}
0 & = & \int_0^1 d\lambda \, \chi_{\lambda}(\im u; \mathbf{q}=0,\mathbf{q}') \nonumber\\
& = & \langle \mathbf{q}=0 | \ln( \kappa_{\ACF}(\im u) )V^{-1} | \mathbf{q}' \rangle \nonumber \\
& = & - \sum_{n=1}^{\infty}\,\frac{1}{n} \, \langle \mathbf{q}=0 
| \left( \tilde{\chi}_{\ACF}(\im u) V \right)^n V^{-1} | \mathbf{q}' \rangle \;.
        \label{eq:effectiveCriterial}
\end{eqnarray}
A sufficient condition for conservation is therefore
\begin{equation}
\tilde{\chi}_{\ACF}(\im u,\mathbf{q}=0,\mathbf{q}') \equiv 0 \; .
\label{eq:chiACFconserve}
\end{equation}

The longitudinal projection of $\epsilon$ allows us to express
\begin{equation}
\tilde{\chi}_{\ACF}(\im u,\mathbf{q},\mathbf{q}') =  
4\pi \mathbf{q}\cdot \mathbf{q}' \langle \mathbf{q} | (1-S)^{-1} S | \mathbf{q} \rangle\,.
\label{eq:chitildeexpand}
\end{equation}
Since $S(z)$ remains finite and free of poles in the upper right quadrant of the complex plane, condition (\ref{eq:chiACFconserve}) is fulfilled.
Conservation of the exchange-correlation hole is therefore inherent to the exchange-correlation description expressed in 
 Eq.~(\ref{eq:FirstExcRecast}) and follows by the principles discussed in Refs.\ \onlinecite{Dobson96p1780}. 

\subsection{vdW-DF in practice}

Functionals designed within the vdW-DF framework do not rely directly on Eq.~(\ref{eq:FirstExcRecast}).
Taking inspiration from the analysis of the
plasmon-based analysis of surface-energy corrections,\cite{Langreth75p1425, Langreth77p2884}
the vdW-DF method splits the total exchange-correlation energy functional
into semilocal and nonlocal contributions,
\begin{equation}
  E_{\rm xc}^{\rm vdWDF}[n] = E_{xc}^{\msl}[n] + E_{c}^{\nl}[n] \, .
\label{eq:ExcGenSplit} 
\end{equation}
Several different exchange functionals have been suggested for the exchange part of $E_{xc}^{\msl}[n]$ as detailed in the main text. Only the LDA part of the correlation is included to avoid double counting semi-local correlation effects. 

The vdW-DF method also considers GGA-type exchange-correlation holes \cite{Rydberg00p6997,Rydberg01p,Dion04p246401,Hyldgaard13p}
 defined by an inner functional
\begin{align}
E_{xc}^{0,i}   + E_{\self} =   \int_0^{\infty} \frac {du}{2\pi} \hbox{Tr} \{ \ln(\epsilon(\im u))\} \, ,
\label{eq:Exc0iguide}
\end{align}
that is suggested by the GGA tradition.\cite{Langreth75p1425, Langreth77p2884,Langreth81p446,Perdew86p8800,Perdew96p3865,Perdew96p16533,Staroverov04p75102} 
A specific functional is selected to describe $E_{xc}^{0,i}$ which in turn introduces a local parameter in $S$. 
Subtracting this term off the one in Eq.~(\ref{eq:ACFexpress}), we obtain the nonlocal correlation energy
\begin{align}
E_{c}^{\nl} &  =\int_0^{\infty} \, \frac{du}{2\pi} \, 
\left[ \hbox{Tr}\ln(\kappa_{\ACF}(\im u)) - \hbox{Tr} \ln(\epsilon(\im u))  \right] \, .
\label{eq:nonlocfunctionaldef}
\end{align}
In the general geometry versions of vdW-DF, this term is further expanded to second order in $S$. 

In general $E^{\rm sl}_{xc} \neq E^{0,i}_{xc}$ and the vdW-DF versions
formally approximate the
exchange-correlation energy
\begin{eqnarray}
E_{xc}[n] & = & E_{xc}^{\vdWDF}[n] + \delta E_{xc}[n] \, ,
\label{eq:ExcSplit}\\[0.7em]
\delta E_{xc} & = &
E_{xc}^{0,i}
-
E_{xc}^{\msl}
\, .
\label{eq:deltaExdef}
\end{eqnarray}
The mismatch is justified by the fact that 
$S$ is designed both with the aim of fulfilling formal constraints as well as to make $E^{\nl}_{c}$ simple to implement. 
Because of its simple form, the inner exchange-correlation hole of vdW-DF does not capture short-range exchange-correlation effects at the same level of sophistication as numerical GGA.\cite{Perdew96p16533,Hyldgaard13p} 
However, this mismatch has the consequence that the automatic exchange-correlation hole conservation secured by starting directly from Eq.~(\ref{eq:FirstExcRecast}) is lost. 
vdW-DF1-cx is designed by using an exchange functional for
$E^{\rm sl}_{xc}$
that makes the semi-local term resemble $E^{0,i}_{xc}$ as closely as feasible. This version largely restores  the automatic conservation  of the  exchange-correlation hole as detailed in Ref.~\onlinecite{Berland13p, Chakarova05p54102}.

\bibliographystyle{apsrev4-1}
\bibliography{new-refs-added-by-KL,vdW-DF-cxES6}
\end{document}